\documentstyle[11pt]{article}
        \def\be{\begin{equation}}
        \def\ee{\end{equation}}

\begin{document}

\begin{center}
{\LARGE{\bf Analysis of Poisson Networks
and Their Relation with Random Cellular Structures }}\\
\vspace{3cm}
V. Karimipour$^{1,2}$ \ \
K. Saeidi$^2$

$^1$
{\it Institute for Studies in Theoretical Physics and Mathematics,\\
P. O. Box 19395-5531, Tehran, Iran}\\

$^2$
{\it Department of Physics, Sharif University of Technology,\\
 P. O. Box  11365-9161, Tehran, Iran} \\
\vspace{2cm}
 Abstract:
\end{center}
We perform a detailed analysis of the statistical properties of Poisson networks
and show that the metric and topological properties of random cellular
structures, can not be derived from simple models of random networks
based on a poisson point distribution [1]. In particular we show that
Lewis and Aboav-Wieare laws are not obeyed in these network.
\newpage
\section {Introduction}

In the past two decades there has been a considerable interest
       in studying a class of non-equilibrium systems known collectively
       as {\it Cellular Structures}[2]. The geometrical and dynamical properties
       of these systems are best displayed in the familiar pattern formed
       by a soap froth confined between two transparent plates . Other
       examples include polycrystalline domains in metals, ceramics,magnetic
        domains and monolayers of fatty acids on surface of water.
       One can also mention other examples from material science, like cracking
       in glazes, fracture and dewetting  of polymer films above
       their glass transition temperature [2].
       Despite the diversity in systems in which cellular structures  are
       formed, numerous experiments have shown that the long time statistical
       behaviour of these systems are  charaterized by  certain
       universal, system independent laws. This means that topological  and
       geometrical constraints influence the properties of these networks
       in a very essential way. For example, the fact that energy is associated
       with the length of the edges  of the cells, immediately
       leads to the conclusion that in the two dimensional  structures all the
       vertices are 3-valent. Combination of this result with the Euler character
        formula shows that the average number of sides per cell,
       $\langle i \rangle$ is equal to 6:
\be
\langle i \rangle=6
\ee
       Similar considerations in 3-dimensions, where energy is associated
       with the area of faces of cells, proves that all vertices are 4-valent
       and that :
\be
   \langle f \rangle = \frac{12}{6-\langle i \rangle}
\ee
       where $\langle f \rangle$ and $\langle i \rangle$ are  the  average
number of faces per cell and the average number of edges per face
 respectivly. Besides these properties which are a direct consequence of
 the Euler character formula, expriments have revealed a number of very
 general, properteis among which the most important are : i) von-Neumann [3],
ii) Lewis [4] and iii) Aboav-Weare law [5,6]. Von-Neumann law refers to the rate
of expansion of a single cell and is accounted for theoritically in a
satisfactory way. It has also been generalized to curved surfaces where
the curvature of the surfaces plays a role both in the dynamic of a single
cell and in it's stability [7]. The Aboav Weare law which is statistical in nature
and refers to the correlation of adjacent cells has been shown to hold in
random 3-valent graphs, viewed as planer feynman graphs of $\phi^3$ theory
and solved by techniques of matrix models [8].\\
However there has been no explanation of the empirical Lewis law which states
that the average area of cells has a linear relation with the number of cells,
for large number of edges.\\
However recently there have been some attempts [1,9] to derive both Lewis law
and Aboav-weare law, from poisson networks, i.e. networks based on a poisson
distribution of horizontal segments between a fixed set of parallel lines.
It has been claimed [1] that in such networks both laws are obeyed.
Also it has been claimed [1] that in one type of the 3-dimensional poison
networks, 3-dimensional analogues of Lewis and Aboav-Weare law are obeyed.
In this paper we study closely these poisson networks and show that the Lewis
and Aboav-Weare law which is derived from these models contain highly
non-linear terms, which is in sharp contrast with experimental data.

The structure of this paper is as follows: in section 2, we introduce the
poisson network in general, such that we can use the result of this section
for both 2-dimensional and 3-dimensional networks. In section 3 we show that
Lewis and Aboav-weare laws are not obeyed in these networks.In section 4 we
show that even if one introduces an extra element of randomness into these
models, namely, non-uniform density distribution, one can not obtain Lewis and
 Aboav-Weare law, although one can obtain better values for the moment of
 distributions of number of edges and faces in 2 and 3-dimensional cellular
 structure. Our coclusions are given in section 5. \\
\section { Poisson networks with uniform density}

The two dimensional poisson networks are generated as follows (figure 1)[1].
One takes a family of parallel lines in the y-direction in the plane .
The distance  between lines does not affect the  topological
properties of the network, although it affectes the geometry . This
distance is taken as uniform and equal to $d_0$. Suppose there are N
columns , $C_{1}, C_2, \ldots, C_{N}$, in the network. The region between two successive
columns is divided into cells.
The division of each column $C_{\alpha}$,is based on a poisson point
distribution in the y-direction. An edge in the x-direction is
taken through each Poisson point (P-point) . In this way a 3-valent
network is generated in the plane which, although different from the
realistic cellular structures, is simple enough to be analysed closely for the
study of various statistical-topological propertis of the network. The
3-dimentional tetravalent network are generated by a similar
 process (figure 2 )[1]. One takes an arbitrary 2-dimensional trivalent network
(base network) on the xy plane (figure 2(a)) and takes vertical planes
(parallel to the z-axis) through each edge  which  divides the 3-dimensional space
into prismatic columns. In each column one considers a Poisson distribution
of uniform density in the z-direction  and divides the
columns into cells by planes perpendicular to the z-axis through each P-point
(figure 2(b)). The height of a cell, $L$, is the distance between
adjacent P-points in the associated distribution (figure 2(c)).The average
cell height is unity.
Consider one of the columns,say $C_{\alpha}$. The Poisson distribution of
$n$ points in a segment of length $L$ is given by :
\be
  p(n)=\frac{L^n}{n!} e^{-L}.
\ee
Cosider a particular set of columns $C_{\alpha}, C_{\beta}, \ldots ,C_{\gamma}$.
The probability that in a given distance $L$, there are $n_{\alpha},n_{\beta},
\ldots ,n_{\gamma}$ segments respectivly in columns $C_{\alpha}, C_{\beta},
\ldots , C_\gamma$, is equal to
\be
\Phi(L, n_\alpha, n_\beta, \ldots , n_\gamma) \equiv
P(n_\alpha) P(n_\beta) \ldots  P(n_\gamma)
\ee
Now consider a reference column which we denote by $C_{\alpha}$. The probability
that there is a cell with height between $L$ and $L+dL$ in this column is
       $e^{-L}dL$.
We denote the columns which are neighbors of this reference column by
$C_{\alpha_1}, C_{\alpha_2},\ldots , C_{\alpha_k}$. The number of these neighboring
columns depend on the dimensionality and type of the network. The joint probability
that there is a cell of height between $L$ and $L+dL$ in the column $C_\alpha$
 such that its neighbors $C_{\alpha_1}, C_{\alpha_2},\ldots , C_{\alpha_k}$
  respectively have $ n_{\alpha_1}, n_{\alpha_2}, \ldots , n_{\alpha_k}$ points in the interval $L$ is :
\be
\Psi_{\alpha}(L, n_{\alpha_1}, n_{\alpha_2}, \ldots , n_{\alpha_k})dL =
P(n_{\alpha_1}) P(n_{\alpha_2}) \ldots  P(n_{\alpha_k}) e^{-L}dL
\ee
clearly this distribution function depends on the individual values of the
variables  $n_{\alpha_1}, n_{\alpha_2}, \ldots , n_{\alpha_k}$, however
in the sequel another probability distribution will be useful. that is
$\Psi(L,I)dL$ were I is defined as $I= n_{\alpha_1}+n_{\alpha_2}+\ldots +
 n_{\alpha_k}$. The joint probability$\Psi(L,I)dL$, is calculated as follows:
 \be
 \Psi_\alpha(L,I)dL= \sum' \Psi_{\alpha}(L,n_{\alpha_1},n_{\alpha_2},
 \ldots,n_{\alpha_k})dL=\frac{(kL)^I}{I!}e^{-(k+1)L}dL.
 \ee
where $ \sum' $ means the sum over all $ n_{\alpha -i } $'s subject to the constraint
their sum be equal to I.
 It's important to note that the distribution $\Psi_\alpha(L,I)$ is obtained
 from a sum of distribution of the form $\Psi_{\alpha}(L,n_{\alpha_1},n_{\alpha_2},
 \ldots,n_{\alpha_k})$ and is not equal to any of them.(compare with eq.(9a)
  of[1]).
  \section { statistical properties of poisson networks}

 { \Large 3.1 Two dimensional networks}

  In two dimensions, every column has two neighbors, hence in eq.(6), $k=2$, and
  the number of edges of a cell (see fig.(1)) in a reference column is equal to :
  \be
  i=I+4
  \ee
  Combining eq.(6) and (7) gives the probability distribution $g(L,i)$ of
  finding {\it i-cells} of hight in the interval $(L, L+dL)$:
  \be
  g(L,i)dL=\frac{(2L)^{(i-4)}}{(i-4)!}e^{-3L}dL.
  \ee
  clearly this distribution function is normalized.
  \be
  \int_{0}^{\infty} \sum_{i=4}^{\infty} g(L,i)dL=1.
  \ee
  From eq.(8) one can obtain the average height of {\it i-cells}:
  \be
  <L>_i=\int_{0}^{\infty} Lg(L,i)dL=\frac{i-3}{3^2}(\frac{2}{3})^{i-4}
  \ee
  This equation, then can be used to calculate the average area of
  {\it i-cells}. For the poisson networks considered in [1], where the width
  of all cells are equal to $d_0$, we find:
  \be
  <A>_i=d_0<L>_i=\frac{d_0(i-3)}{3^2}(\frac{2}{3})^{i-4}
  \ee
  Clearly this relation, due to it's non-linear term, is in sharp contrast
  with Lewis law.
  \\
  From eq.(8), one can obtain the total probability distribution of {\it
  i-cells}$\ \  g(i)$:
  \be
  g(i)=\int_{0}^{\infty} g(L,i)dL=\frac{1}{2}(\frac{2}{3})^{i-3}
  \ee
  One can also define a generating function :
  \be
  G(q)=<e^{iq}>\equiv\sum_{i=4}^{\infty} g(i)e^{iq}
  \ee
  A simple calculation shows that:
  \be
  G(q)=\frac{e^{4q}}{3-2e^q}
  \ee
  or
  \be
  lnG(q)=4q-ln(3-2e^q)
  \ee
  From which one obtains by successive differentiation various connected
  moment of the ditribution:
  \be
  <i>=6
  \ee
  \be
  <i^2>-<i>=6
  \ee
  \be
  <i^3>-3<i^2><i>+2<i>^3=30
  \ee    \newpage
{\Large3.2.  three-dimensional networks.}

Consider a 3-dimensional network and a column whose base is an {\it i-cell.}
In eq.(6), we should now equate k with i and, relate I with the number of faces as follows:
\be
I+2+i=F.
\ee
The probability distribution of {\it F-cells} (cells with F-faces) of height
between$L$ and $L+dL$, is now obtained as:
\be
g_i(L,F)dL=\frac{(iL)^{F-i-2}}{(F-i-2)!}e^{-(i+1)L}dL
\ee
From which one can obtain the average height of {\it F-cells} whose base are
{\it i-cells}.
  \be
<L>_{i,F}=\int_{0}^{\infty} Lg_i(L,F)dL=\frac{F-i-1}{(i+1)^2}(\frac{i}{i+1})^{F-i-2}
  \ee
The average height of {\it F-cells} with any base is given by
\be
<L>_F=\sum_{i=4}^{\infty} <L>_{F,i}g(i)
\ee
where $g(i)$ is the probability distribution of {\it i-cells} in the base. As
an approximaion one can set $i=<i>=6$ in (21) and obtain:
\be
<L>_F=\frac{F-7}{7^2}(\frac{6}{7})^{F-8}
\ee
Compare with eq.(36) of [1]. The average volume of {\it F-cells} will than be:
\be
<v>_F=S\frac{F-7}{7^2}(\frac{6}{7})^{F-8}
\ee
where $S$ is the average area of cells in the base.

Here again, it's seen that there is a pronounced non-linear term, which however
is not as worse as the two dimensional case. From eq.(20) one can also find the
total ditribution of {\it F-cells}  with {\it i-cell} bases:
 \be
g_i(F)=\frac{1}{i+1}(\frac{i}{i+1})^{F-i-2}
  \ee
From this formula various monents can be calculated:
\be
<F>_i=2i+2
\ee
\be
<F^2>=58+5i^2
\ee
and
\be
<F>=2<i>+2=14
\ee
\be
<F^2>=58+5<i^2>=268
\ee
In the remaining part of this section we review breifly the basic steps of the
analysis of ref.[1] and show that if one uses the correct form of the
average $<L>_i$ as given by eq.(10), then for a poisson network, Aboav-Weare
law will not be obeyed either, even approximatly.\\
Consider an {\it i-cell} $a$ of length $L$ in a two dimensional poisson
network(figure 3). This cell is in column $a$ and has two neighbors in this
column, called $a'$ and $a''$.\\
There are two adjacent neighbors $b_1$ and $b_2$, which respectively
distribute $n_1$ and $n_2$ points inside the cell $a$. The number of
sides of $a$ is then equal to $L$
\be
i=n_1+n_2+4
\ee
The total number of sides of the cells in column $b_1$ adjacent to $a$ is
\be
J_1=4(n_1+1)+(m_1+1)+(m'_1+1)+m''_1
\ee
where the meaning of the numbers $ m_1 , m_1' $ and $ m_1" $ are specified in
fig. (3).
Clearly $im_i$ which is the total number of sides of the cells adjacent to
$a$ is:
\be
im_i=12+J_1+J_2.
\ee
Where the definition of $J_2$ is similar to that of $J_1$. For the average
value of $ m_1, m'_1  $ and  $ m''_1 $ we use:
\be
<m_1+m'_1>=\frac{<L_c-L_a>}{1}=<L_c-L_a>
\ee
\be
<m''>=\frac{<L_{c_b}>}{1}=<L_{c_b}-L_a>+<L_a>
\ee
Where the segment $L_{c_b}$ in column $b_1$, which is the smallest segment
containig $L_a$, is called the cvovering  lenght of $l_a$ [1].
With the probability distribution found in [1] for $L_c$
it is shown that the average $<L_c-L_a>=2$, it then follows that:
\be
<im_i>=12+4(n_1+n_2)+<m_1+m'_1>+<m_2+m'_2>+<m''_1>+<m''_2>+4
\ee
The number $12$ comes from the average of sides of $a'$ and $a''$.
Combining eq.(33), (34) and (35) one finds that:
\be
<im_{i}>=16+4i+2<L>_i
\ee
Combining eq.(36) with (10) one obtains
\be
<im_{i}>=16+4i+\frac{2(i-3)}{3^2}(\frac{2}{3})^{i-4}
\ee
which shows that Aboav-Weare law is not valid for such poisson networks.
Clearly Poisson networks by construction do not have enough randomness to
be good models of cellular structures, as far as geometrical propertis
(i.e. Lewis law) are concerned.One may even  try to introduce more randomness,
 (i.e. non- uniform point distribution in different columns) to see if the
 non-linear terms shown in eq.(37) can be smoothed out. In the next section
 we examine breifly this possibility. The result is that although one can
 use this extra degree of freedom to obtain better agreement with exprimental
 data for various moments of number of sides(i.e. $<i^2>$, etc.), again the Lewis and
 Aboav-weare law will not be valid.\\
 \section {Poisson networks with non-uniform density}

 A little modification of the previous formulas is necessery
 to study the propertis of a network where the density of points (the average
 distance between points to be denoted by $\lambda$, in the following)
 in different columns are different. Instead of (1) we will have:
 \be
 P_\alpha(n)=(\frac{L}{\lambda_\alpha})^{n}\frac{1}{n!}
 e^{-\frac{L}{\lambda_\alpha}}
 \ee
Instead of (4) we will have:
\be
\Phi_{0}(L, n_1, n_2, \ldots, n_k)=P_{1}(n_1)P_{2}(n_2)\ldots
P_{k}(n_k)=\prod_{\alpha=1}^{k} (\frac{L}{\lambda_\alpha})^{n_\alpha}
\frac{1}{n_\alpha!}e^{-\frac{L}{\lambda_\alpha}}
\ee
defining $\Psi(L,I)dL$ as before, we obtain:
\be
\Psi_0(L,I)=\sum'_{n_1+n_2+\ldots+n_k=I} \Phi(L, n_1, n_2, \ldots, n_k)
\frac{1}{\lambda_0}e^{-\frac{1}{\lambda_0}}dL=
\frac{L^{I}(\xi_0-\frac{1}{\lambda_0})^{I}}{\lambda_{0}{I!}}e^{-L{\xi_0}}
\ee
where $\xi_0=\frac{1}{\lambda_0}+\frac{1}{\lambda_1}+\frac{1}{\lambda_2}+
\ldots+\frac{1}{\lambda_k}$.
In a two dimensional network every column say $C_{n}$ has two neighbors,
$C_{n-1}$ and $C_{n+1}$. The number of sides of {\it i}-cell(a cell with {\it i}-edges) in column n, is
$i=4+I$, where $I=n_{n-1}+n_{n+1}$ and $n_{n-1}+n_{n+1}$ are the number of
vertices contributed by the cells in adjacent columns. Then for the distribution
 of {\it i-cells} of length between $(L $ and $ L+dL)$ in column $C_{n}$, we obtain:
\be
g_n(L,i)= \frac{L^{i-4}(\xi_n-\frac{1}{\lambda_n})^{i-4}}{\lambda_n{(i-4)}!}
e^{-L\xi_n}
\ee
where $\xi_n=\frac{1}{\lambda_{n-1}}+\frac{1}{\lambda_{n}}+\frac{1}{\lambda_{n+1}}$
We will also obtain, the distribution of {\it i-cells} in column $C_n$:
\be
g_n(i)=\int_{0}^{\infty} g_n(L,i)dL= \frac{1}{\lambda_n\xi_n}
(1-\frac{1}{\lambda_n\xi_n})^{i-4}
\ee
Clearly this distribution function is normalized.
We can now evaluate various moments of the distribiution $g_{n}(i)$. In all the
 following calculations, we use the following formulas for geometric sum:
 \be
 \sum_{k=0}^{\infty} x^{k} = \frac{1}{1-x}
 \ee
 \be
 \sum_{k=0}^{\infty} kx^{k} = \frac{x}{(1-x)^{2}}
 \ee
 \be
\sum_{k=0}^{\infty} k^{2}x^{k} = \frac{x (1+x)}{(1-x)^{3}}
\ee
A simple calculation shows that:
\be
<i>_{n}=\sum_{i=4}^{\infty} ig_{n}(i)=3+\lambda_{n}\xi_{n}=4+\lambda_{n}
(\frac{1}{\lambda_{n-1}}+\frac{1}{\lambda_{n+1}})
\ee
where $<i>_{n}$ is the average number of {\it i-cells} in column n. For a uniform
density, this number turns out to be equal to 6. Another quantity of interest is $<i^2>_{n}
=\sum_{i=4}^{\infty} i^2g_{n}(i)$, which turns out to be :
\be
<i^2>_n=2(\lambda_{n}\xi_{n})^2+5\lambda_{n}\xi_{n}+9
\ee
The variance turns out to be equal to:
\be
<i^2>_n-<i>_n^2=(\lambda_{n}\xi_{n})(\lambda_{n}\xi_{n}-1)
\ee
In order to find the moments $<i>$ and $<i^2>$ in the whole lattice and not in a
particular column n, we replace the average over
the cells in a network, by an average over the densities in an ensemble of
networks. Thus from (46) and (47) we have :
\be
<i>=4+<\lambda_{n}(\frac{1}{\lambda_{n-1}}+\frac{1}{\lambda_{n+1}})>
\ee
\be
<i^2>=2<(\lambda_{n}\xi_{n})^2>+5<\lambda_{n}\xi_{n}>+9
\ee
where the averages on the right hand side are performed
with a suitable distribution $$P(\lambda_{1}, \lambda_{2}, \ldots , \lambda_{N}).$$
For simplicity in the following we assume that this distribution, is a symmetric
distribution, i.e:
\be
P(\lambda_{1}, \lambda_{2}, \ldots , \lambda_{N})=
 P(\lambda_{1'}, \lambda_{2'}, \ldots , \lambda_{N'})
 \ee
  where the primed indices are a permutation of the unprimed ones.
 Clearly this minor restriction allows our conclusions to be valid for a very
 large class of probability distributions.
 For a symmetric ditribution we can simplify (49) and obtain:
 \be
<i>=6
\ee
 as expected. Note that for a uniform distribution, we obtain
 \be
<i^2>=42 \hspace{1cm}<i^2>-<i>^2=6
\ee
 in agreement with [1].
 From (41) we also obtain the average height of {\it i-cells} in the nth column:
 \be
 <L>_{n,i}=\int_{0}^{\infty} Lg_n(L,i)dL=\frac{i-3}{\lambda_n\xi_n^2}
 (1-\frac{1}{\lambda_N\xi_n})^{i-4}
 \ee
 Again one sees that the non-linear behaviour persists in this average.\\
\section {Three-Dimentional Networks}

Consider a 3-dimentional laminated poisson network, based on an arbitrary
2-dimentional network, and a perticular {\it i-cell} is the base. We label
the column with this base by $C_{i,0}$. The number of faces in a
 cell in the column
above this {\it i-cell} is
\be
F = i + 2 +J
\ee
where $J=m_1+m_2+ \ldots + m_i$, and $m_k$ is the number of additional lateral
 faces which results from the cells in the adjacent $k-$column (fig. 4).
Let $f_i(J)$ be the fraction of the $F$ cells in column $C_{i,0}$.
In formula (40), one should now replace $k$ by i, and I by J to obtain:
\be
\Psi_0(L,F,i)=\frac{L^{F-i-2}(\xi_0-\frac{1}{\lambda_0})^{F-i-2}}{\lambda_0J!}
e^{-L\xi_0}
\ee
where $\xi_0=\frac{1}{\lambda_0}+\frac{1}{\lambda_1}+\frac{1}{\lambda_2}+
\ldots+\frac{1}{\lambda_i}$. Here $\lambda_0$ is the density of P-point distribution in the column
$C_{i,0}$ and $\lambda_\alpha$ ($\alpha=1,2,\ldots,i$) are the density of
 P-point distribution in the adjacent columns. $\Psi_0(L,F,i)dL$ is the
probability of finding an {\it F-cell} in column $C_{i,0}$, whose height is
between $L$ and $L+dL$. Integrating over $L$, one obtains
\be
\Psi_0(F,i)=\frac{1}{\lambda_0\xi_0}(1-\frac{1}{\lambda_0\xi_0})^{F-i-2}
\ee
This is the probability of finding an {\it F-cell} in column $C_{i,0}$.
One can now find various moments of this distribution>:
\be
<F>_{i,0}=\lambda_0\xi_0+i+1
\ee
\be
<F^2>_{i,0}= (i+1)^2+ (2i+1)(\lambda_0\xi_0)+2(\lambda_0\xi_0)^2
\ee
Note that for uniform density one has :$\lambda_0\xi_0=i+1$
and the above formula reduce to
\be
<F>_i=2i+2
\ee
\be
<F^2>_i=(i+1)(5i+4)
\ee
in agreement with[].
By the same averaging procedure as in the two dimensional case one can obtain
\be
<F>_i=\lambda_0\xi_0+i+1
\ee
\be
<F^2>_i= (i+1)^2+ (2i+1)(\lambda_0\xi_0)+2(\lambda_0\xi_0)^2
\ee
The later averages are performed over the density distributions.\\
\section {Conclusion}

Our main conclusion is that Lewis and Aboav-Weare laws do not apply to
poisson networks.This is due to the rather rigid geometrical structure
of these networks. We have shown that by more randomizing these networks,
(i.e. in the sence of random density distribution in different
columns), one can not still obtain the above mentioned laws. In fact the only feature
that these networks have in common with real cellular structures is that they
are 3-valent, and the main property of real systems in which  all the angles
around any vertex are equal is oboviously absent in these networks.It seems
that the Lewis and to some extant Aboav-Weare law which are inherently
geometrical in nature will not show up in those random models (i.e. Poisson)
which neglect this very specific property of cellular structures.
\newpage
{\large \bf References}
\begin{enumerate}
\item  Fortes M.A.; J. Phys. A : Math. Gen. {\bf 28} (1995)1055-1068
\item  For a recent review see : J. Stavans ; " The Evolution of Cellular Structures "
Rep. Prog. Phys. {\bf 56}(1993) 733-789, and references therein.
\item  von Neumann J. " Metal Interfaces " ( Cleveland, OH: American Society for Metals)
p. 108
\item Lewis F. T. Anat. Rec. {\bf 38} (1928) 241
\item Aboav D. A. :Metallography {\bf 3} (1970) 383-90
\item Aboav D. A. :Metallography {\bf 13} (1980) 43-58
\item Avron J. E. and Levine D. : Phys. Rev. Letts. {\bf 69} (1992) 208
\item Godreche C., Kostov I. and Yekutili I. : Phys. Rev. Letts. {\bf 69} (1992) 2674
\item Fortes M. A. Phil. Mag. Lett. {\bf 68} (1993) 69-71

\end{enumerate}
\end {document}